\documentclass[12pt]{article}
\usepackage[top=1.0in, bottom=1.0in, left=1.12in, right=1.12in]{geometry}
\usepackage{graphicx} %
\usepackage{caption}
\usepackage{subcaption}
\usepackage{algorithm}
\usepackage{algorithmic}
\usepackage{hyperref}
\usepackage{cleveref}
\usepackage{amsmath, amssymb, amsthm}

\usepackage{natbib}

\usepackage{color}
\usepackage{booktabs}

\usepackage{setspace}
\renewcommand\footnotemark{}

\begin{document}
\title{Differential Expression Analysis of Dynamical Sequencing Count Data with a Gamma Markov Chain}
\author{%
 Ehsan Hajiramezanali\,$^{\text{1}}$, Siamak Zamani Dadaneh\,$^{\text{1,2}}$,
Paul de Figueiredo\,$^{\text{3,4}}$,\\
Sing-Hoi Sze\,$^{\text{5}}$,
Mingyuan Zhou\,$^{\text{6}}$, and Xiaoning Qian\,$^{\text{1,2,*}}$
\thanks{\newline
$^{\text{\sf 1}}$Department of Electrical and Computer Engineering, Texas A\&M University, College Station, TX 77843, USA, \newline
$^{\text{\sf 2}}$Center for Bioinformatics \& Genomic Systems Engineering, Texas A\&M University, College Station, TX 77843, USA, \newline
$^{\text{\sf 3}}$Department of Microbial Pathogenesis and Immunology, Texas A\&M Health Science Center, College Station, TX 77843, USA, \newline
$^{\text{\sf 4}}$Department of Veterinary Pathobiology, and Norman Borlaug Center, Texas A\&M University, 77843, USA, \newline
$^{\text{\sf 5}}$Department of Computer Science and Engineering, and Department of Biochemistry \& Biophysics, Texas A\&M University, College Station, TX 77843, USA, \newline
$^{\text{\sf 6}}$ Department of Information, Risk, and Operations Management (IROM), The University of Texas at Austin, Austin, TX 78712, USA.\newline
$^{\text{\sf *}}$Correspondence should be addressed to Xiaoning Qian (\texttt{xqian@ece.tamu.edu}).
}
}

\begin{spacing}{1.25}
\maketitle

\begin{abstract}
  Next-generation sequencing (NGS) to profile temporal changes in living systems is gaining more attention for deriving better insights into the underlying biological mechanisms compared to traditional static sequencing experiments. Nonetheless, the majority of existing statistical tools for analyzing NGS data lack the capability of exploiting the richer information embedded in temporal data. Several recent tools have been developed to analyze such data but they typically impose strict model assumptions, such as smoothness on gene expression dynamic changes. To capture a broader range of gene expression dynamic patterns, we develop the gamma Markov negative binomial (GMNB) model that integrates a gamma Markov chain into a negative binomial distribution model, allowing flexible temporal variation in NGS count data. Using Bayes factors, GMNB enables more powerful temporal gene differential expression analysis across different phenotypes or treatment conditions. In addition, it naturally handles the heterogeneity of sequencing depth in different samples, removing the need for ad-hoc normalization. Efficient Gibbs sampling inference of the GMNB model parameters is achieved by exploiting novel data augmentation techniques. Extensive experiments on both simulated and real-world RNA-seq data show that GMNB outperforms existing methods in both receiver operating characteristic (ROC) and precision-recall (PR) curves of differential expression analysis results.

\vspace{4mm}
\noindent%
\end{abstract}
\end{spacing}

\section{Introduction}

Advances in next-generation sequencing (NGS) techniques have enabled researchers to produce
millions of relatively short reads for genome-scale bioinformatics research \citep{encode2012integrated,wang2009rna,mortazavi2008mapping}. Transcriptome analyses,
including gene expression profiling and transcript quantification through RNA-sequencing (RNA-seq), can help better understand biological processes of interest. RNA-seq count data are highly over-dispersed with large dynamic ranges \citep{anders2015htseq}. A large number of statistical tools have been developed for differential gene expression analysis of RNA-seq data \citep{anders2010differential,dadaneh2017bnp,robinson2010edger,love2014moderated,law2014voom,hardcastle2010bayseq,leng2013ebseq}, which mostly have adopted the negative binomial (NB) distribution to account for over-dispersion as well as high uncertainty inherent in RNA-seq data due to the small number of replicate samples in typical differential expression experiments~\citep{love2014moderated}.

Living systems are complex and dynamic. There has been significant interest in analyzing temporal RNA-seq count data \citep{bar2012studying}. For example, in cell biology or drug discovery research, monitoring molecular expression changes in response to specific stimuli can help better understand cellular mechanisms at the transcriptional and post-transcriptional regulatory levels under different conditions. One important task is to identify the genes that are differentially expressed over time across different conditions, which is more challenging compared to static RNA-seq data analysis due to potential temporal dependencies \citep{lienau2009insight}.

Recently, several dynamic differential RNA-seq analysis methods have been developed to better capture temporal dependency. For example, EBSeq-HMM \citep{leng2015ebseq} takes an empirical Bayesian mixture modeling approach to compare the expression change across consecutive time points to identify  genes that display significant transcription changes over time under one treatment condition. Across different conditions, it is desirable to identify genes that have different dynamic patterns. For this purpose, next-maSigPro \citep{nueda2014next} has extended a generalized linear model (GLM) \citep{mccullagh1984generalized} based dynamic differential expression analysis for microarray data from multiple time points to analyze temporal RNA-seq data. However, modeling RNA-seq counts by real values may lead to information loss and GLM may not be able to capture complicated dynamic changes in expression. An autoregressive time-lagged $AR(1)$ model with Markov Chain Monte Carlo (MCMC) inference~\citep{oh2013time} has also been proposed to identify genes with different temporal expression changes. But the posterior estimates of model parameters through Metropolis-Hastings inference lead to high computational complexity. DyNB \citep{aijo2014methods} has been proposed recently to model the temporal RNA-seq counts by NB distributions with their temporal expected values modeled by non-parametric Gaussian Processes (GP). DyNB can detect the genes with differential dynamic patterns that static differential expression analysis, which consider individual time points, fail to discover. In addition to high computational complexity due to MCMC inference \citep{spies2015dynamics,sun2016statistical}, DyNB may fail to model potential abrupt expression changes due to its inherent smoothness assumptions~\citep{rasmussen2006gaussian}.

We present a new dynamic differential expression analysis method for temporal RNA-seq data, GMNB (gamma Markov negative binomial), which is a hierarchical model to introduce a gamma Markov chain \citep{acharya2015nonparametric,schein2016poisson} to model the potential dynamic transitions of the model parameters in NB distributions. With this new model for temporal RNA-seq data and an efficient inference algorithm, GMNB is expected to provide the following advantages over existing methods: 1) GMNB can model more general dynamic expression patterns than DyNB, especially for abrupt expression changes across consecutive time points; 2) The closed-form Gibbs sampling can be derived to infer the model parameters in GMNB, which is computationally more efficient than the existing methods; 3) For dynamic differential expression, genes are ranked based on the Bayes factor (BF), which is very general especially when considering differential expression under multiple factors; 4) Last but not least, GMNB avoids the normalization preprocessing step due to the explicit modeling of the sequencing depth in NB distributions, as described in \citet{dadaneh2017bnp},  %
and we expect similar superior performance of GMNB compared to existing methods requiring such heuristic preprocessing steps.

The remainder of the paper is organized as follows. Section 2 introduces the GMNB model, inference algorithm, and dynamic differential expression analysis. Section 3 compares the experimental
results from both synthetic and real-world benchmark data using GMNB and other state-of-the-art dynamic differential expression methods for temporal RNA-seq data. We conclude the paper in Section 4.

\section{Methods}

\subsection{Notation}

Throughout this paper, we use the NB distribution to model RNA-seq read counts. We parameterize a NB random variable as $n \sim \text{NB}(r,p)$, where $r$ is the nonnegative dispersion and $p$ is the probability parameter. The probability mass function (pmf) of $n$ is expressed as $f_N(n)=\frac{\Gamma(n+r)}{n!\Gamma(r)}p^n(1-p)^r$, where $\Gamma(\cdot)$ is the gamma function. The NB random variable $n \sim \text{NB}(r,p)$ can be generated from a compound Poisson distribution:  
\begin{equation}
n = \sum_{t=1}^{\ell} u_t, \;\; u_t \sim \text{Log}(p), \;\; \ell \sim \text{Pois}(-r\ln (1-p)), \nonumber
\end{equation}
where $u \sim \text{Log}(p)$ corresponds to the logarithmic random variable \citep{johnson2005univariate}, with the pmf $f_U(u) = -\frac{p^u}{u\ln(1-p)}$, $u=1,2,...$. As shown in \citet{zhou2015negative}, given $n$ and $r$, the distribution of $\ell$ is a Chinese Restaurant Table (CRT) distribution, $(\ell | n,r) \sim \text{CRT}(n,r)$, a random variable from which can be generated as %
$\ell = \sum_{t=1}^{n} b_t$, with $b_t \sim \text{Bernoulli}(\frac{r}{r+t-1})$.

\subsection{GMNB model}

We model the dynamic gene expression changes in a temporal RNA-seq dataset by constructing a Markov chain where the expression of a gene at time $t$ only depends on that of time $t-1$. Specifically, for the RNA-seq reads mapped to gene $k$ in a given sample $j$ under different conditions, the read count at time $t$ follows:
\begin{equation} \label{eq:model}
    n_{kj}^{(t)} \sim \text{NB}(r_k^{(t)}, p_j^{(t)}),
\end{equation}
where to impose the dependence between consecutive time points, we model the dispersion parameters dynamically by introducing a gamma Markov chain, in which $r_k^{(t)}$ is distributed according to:
\begin{equation}
    \label{eq:gammamodel}
    r_k^{(t)} \sim \text{Gamma}(r_k^{(t-1)}, \frac{1}{c_k}). 
\end{equation}

As previously shown in \citet{dadaneh2017bnp}, the probability parameter $p_j^{(t)}$ accounts for the effect of varying sequencing depth of sample $j$ at time point $t$. More precisely, the expected expression of gene $k$ in sample $j$ and time $t$ is $r_k^{(t)} \frac{p_j^{(t)}}{1-p_j^{(t)}}$, and hence the dispersion parameter $r_k^{(t)}$ can be viewed as the true abundance of gene $k$ at time $t$, after removing the effects of sequencing depth. Thus the differential expression analysis of temporal RNA-seq data can be performed without any normalization preprocessing steps.

Note that the scale parameter ${1}/{c_k}$ of the Gamma distribution in (\ref{eq:gammamodel}) is shared between different time points, thereby making statistical inference more robust by borrowing information from various samples at multiple time points. To complete the model we sample the dispersion parameter at the first time point as $r_k^{(0)} \sim \text{Gamma}(e^{(0)}, \frac{1}{f_0})$, and use conjugate priors as  $c_k \sim \text{Gamma}(c_0, \frac{1}{d_0})$ and $p_j^{(t)} \sim \text{Beta}(a_0, b_0)$. %

In addition to the flexibility of modeling temporal RNA-seq data, this GMNB model enables an efficient  inference procedure by taking advantage of unique data augmentation and marginalization techniques for the NB distribution \citep{zhou2015negative}, as described in detail below.

\subsection{Gibbs sampling inference}

By exploiting novel data augmentation techniques in \citet{zhou2015negative}, we implement an efficient Gibbs sampling algorithm with closed-form updating steps. More specifically, we infer the dispersion parameter of the NB distribution by first drawing latent random counts from the CRT distribution, and then update the dispersion by employing the gamma-Poisson conjugacy. Furthermore, due to the Markovian construction of the model, it is necessary to consider both backward and forward flow of information for the inference of $r_k^{(t)}$. First, in the backward stage, starting from the last time point $t = T$, we draw two sets of auxiliary random variables as 
\begin{eqnarray}
  l_{kj}^{(t)} &\sim& \text{CRT} (n_{kj}^{(t)}, r_k^{(t)}) \nonumber\\
  l_{k.}^{(t)} &=& \sum_j l_{kj}^{(t)} \nonumber\\
  u_{k}^{(t-1)(t)} &\sim& \text{CRT} (u_{k}^{(t)(t+1)}+l_{k.}^{(t)}, r_k^{(t-1)}),
\end{eqnarray}
for $t=T,T-1,\ldots,1$. %
For the last time point, we assume $u_{k}^{(T)(T+1)}=0$. Next, in the forward stage of Gibbs sampling, we sample $r_{k}^{(t)}$ starting from $t = 0$ to $t = T$ as
\begin{equation}
  (r_{k}^{(t)} | - ) \sim \text{Gamma} \big( r_{k}^{(t-1)} + u_{k}^{(t)(t+1)} + l_{k.}^{(t)}, \frac{1}{\theta_{k}^{(t)}} \big),
\end{equation}
where $\theta_{k}^{(t)} = c_k - \sum_j \ln(1-p_j^{(t)}) - \ln{(1-q_k^{(t)})}$. For $t=0,...,T-1$,  $q_{k}^{(t)}$ is defined as
\begin{equation}
q_{k}^{(t)} = \frac{- \sum_j \ln(1-p_j^{(t+1)})-\ln(1-q_{k}^{(t+1)})}{c_k - \sum_j \ln(1-p_j^{(t+1)})-\ln(1-q_{k}^{(t+1)})},  
\end{equation}
and $q_{k}^{(T)}=0$. Finally, by taking advantage of %
conjugate priors, 
in each iteration of Gibbs sampling, $c_k$ and $p_j^{(t)}$ can be drawn as
\begin{eqnarray}
  (c_k | - ) &\sim& \text{Gamma}(e_0 + \sum_{t=0}^{T-1} r_{k}^{(t)}, 1 / (f_0 + \sum_{t=1}^{T} r_k^{(t)})), \nonumber\\
  (p_j^{(t)} | - ) &\sim& \text{Beta}(a_0 + \sum_k n_{kj}^{(t)}, b_0 + \sum_k r_k^{(t)}).
\end{eqnarray}

The efficient augmentation technique employed in our Gibbs sampling inference removes the need for specifying a suitable proposal distribution, as in the Metropolis-Hastings inference of both DyNB~\citep{aijo2014methods} and NB-AR(1) methods~\citep{oh2013time}. Our experiments in the next section demonstrate that the Gibbs sampling algorithm of GMNB has fast convergence.

\subsection{Dynamic differential expression using Bayes factors}
\label{sec:bf}

The main goal of differential expression analysis is to identify the genes whose expressions demonstrate significant variations across conditions. In the classic static RNA-seq data analysis, this goal is usually obtained via the comparison of expression averages across groups. In dynamic RNA-seq measurement settings, however, this task becomes more challenging as any change of temporal expression patterns between groups may reflect interesting biological mechanisms. Hence, as in \citet{aijo2014methods}, we adopt the Bayes Factor (BF) as a measure that exploits information collectively from all time points to detect the genes with significant variations in temporal expression patterns across conditions. 

To compute the Bayes Factor, we first consider the null hypothesis $\mbox{H}_0$ that the genes are not differentially expressed across conditions, and thus the same set of parameters govern the temporal gene expressions. In this case, we aggregate the counts $\mathcal{D}$ of both experimental conditions to fit the GMNB model $\mbox{M}_0$. On the other hand, under the alternative hypothesis $\mbox{H}_1$, the differentially expressed genes possess different model parameters in each group. Hence, GMNB models $\mbox{M}_1$ and $\mbox{M}_2$ are independently fitted to the counts in conditions 1 ($\mathcal{D}_1$) and 2 ($\mathcal{D}_2$), respectively. Then, the BF can be calculated~as
\begin{eqnarray}
	\mbox{BF} &=& \frac{P(\mathcal{D} | \mbox{H}_1)}{P(\mathcal{D} | \mbox{H}_0)} = \frac{P(\mathcal{D}_1 | \mbox{M}_1) P(\mathcal{D}_2 | \mbox{M}_2)}{P(\mathcal{D} | \mbox{M}_0)}, \nonumber
\end{eqnarray}
where we have assumed equal prior probabilities for both hypotheses. The BF computation requires marginalizing out model parameters, which we conduct through Monte Carlo %
integration using posterior samples collected in the Gibbs sampling procedure. %

\section{Experimental Results}

We evaluate the proposed GMNB model and compare its performance on both synthetic and real-world temporal RNA-seq data with DyNB \citep{aijo2014methods}. We also consider DESeq2 \citep{love2014moderated}, which is a popular tool for differential expression analysis, however, not specifically designed for temporal RNA-seq data. We first consider synthetic RNA-seq data generated by different temporal models, and show that GMNB consistently provides outstanding performance in terms of the area under the curves (AUCs) of receiver operating characteristic (ROC) and precision-recall (PR) curves. Furthermore, we present two case studies on human Th17 cell differentiation \citep{tuomela2016comparative,chan2016subpopulation,aijo2014methods}, and explain the biomedical implications based on differential expression analysis over time by GMNB.

Throughout the experimental studies for synthetic and real-world data, for GMNB, in each run of Gibbs sampling inference $1000$ MCMC samples of parameters are collected after $1000$ burn-in iterations. We use the collected MCMC samples to calculate the BF for each gene as explained in Section~\ref{sec:bf}, and rank the genes according to these BFs. For DyNB, we follow the settings provided in \citet{aijo2014methods} and rank the genes using the computed BFs. We consider three different setups for differential expression analysis of temporal RNA-seq data using DESeq2. In the first setup, denoted by DESeq2-GLM in the experiments, time information is incorporated as a covariate of the generalized linear model in DESeq2 in differential expression analysis to determine temporal data in one model. In the second and third setups, we apply DESeq2 to the data at different time points independently, and use the average and minimum computed p-values from the respective differential expression analyses as an overall measure of differential expression across conditions, denoted by DESeq2-avg and DESeq2-min in the experiments, respectively.

It is worth mentioning that the use of an efficient closed-form Gibbs sampling makes GMNB, on average, 10 times faster than DyNB for both simulated and real-world temporal RNA-seq datasets by reducing the number of iterations required to converge. This is due to the low acceptance rate of the Metropolis-Hastings step of DyNB inference. Thus, to ensure the convergence of its MCMC inference, we consider performing $100,000$ iterations in DyNB for each dataset. On the other hand, our experiments show that as few as $2000$ iterations %
are sufficient for the proposed Gibbs sampling algorithm of GMNB.
%

\subsection{Synthetic data}

We first perform a comprehensive evaluation of GMNB with the synthetic data generated under different temporal RNA-seq models. More precisely, we simulat the data under the following three different setups: the proposed GMNB generative model, the DyNB generative model \citep{aijo2014methods}, and the auto-regressive (AR) based procedure \citep{oh2013time}. In all setups 10\% of genes are randomly set to be truly differentially expressed, with the procedure described in detail for each setup in the following subsections. For each specific generative model, we change the corresponding model parameters to ensure that the expected expression changes of truly differentially expressed genes are different across two conditions.
The impact of sequencing depth variation is simulated by drawing the corresponding size factors from the interval $[0.8,1.2]$ uniformly at random.

\subsubsection{Comparison based on GMNB generative model}

In the first simulation study, we generate the synthetic RNA-seq count data for $1000$ genes under two conditions according to the GMNB model (\ref{eq:model}) with the gamma-Markov temporal dependencies (\ref{eq:gammamodel}) between dispersion parameters. The gene-wise scale parameters $c_k$ are drawn from the uniform distribution in the interval $[0.8,2]$. To simulate 10\% differentially expressed genes, the scale parameter in the second condition are modified to be $c_k + b$, where
$$b =
  \begin{cases}
    0.02       & \quad \text{if } c_k < 1\\
    -0.02  & \quad \text{if } c_k \geq 1
  \end{cases}
$$ determines the significance of differential expression across conditions. The dispersion parameter at the initial time point, $r_k^{(0)}$, is generated for both conditions according to $\mbox{Gamma}(e_0,10)$ where $e_0=\mbox{Uniform}(30,50)$. To simulate the effect of potential varying sequencing depths, the size factors are drawn uniformly at random from the interval $[0.8, 1.2]$. For each condition and each time point, 4 replicates are generated based on the explained procedure.

\begin{figure}[ht]
\begin{center}
\includegraphics[width=.85 \textwidth]{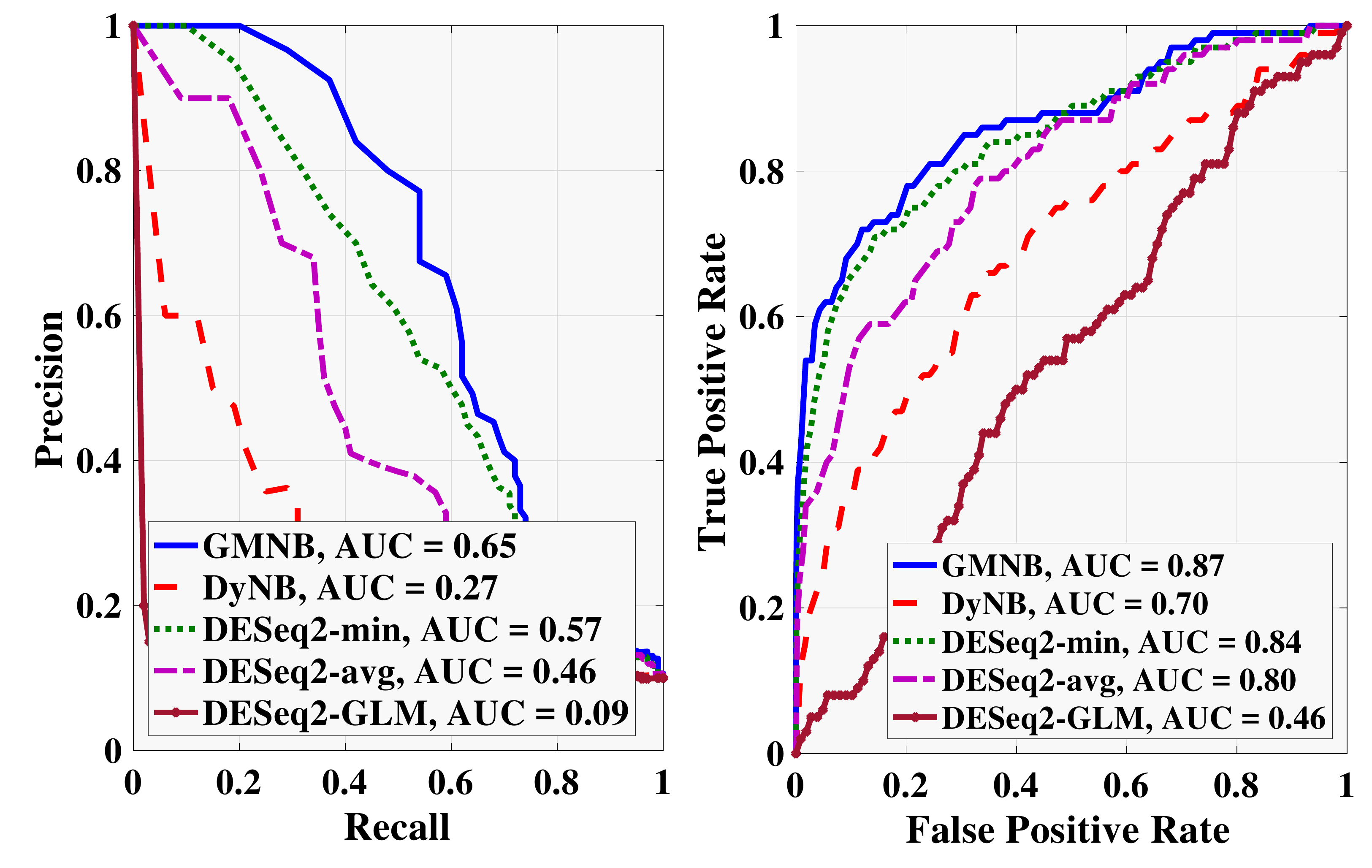}
\caption{{\bf Left column:} PR Curve, {\bf Right column:} ROC Curve. Performance comparison of different methods for differential gene expression over time based on the GMNB generative model. AUCs are given in the corresponding legends in the plots.}
\label{gmnb}
\end{center}
\end{figure}

Figure~\ref{gmnb} illustrates the performance of different methods evaluated based on the simulated data. The proposed GMNB model outperforms the other methods with a significant margin for both ROC and PR curves. The AUCs of both curves are also significantly higher than those by the other methods (in the legends of Figure~\ref{gmnb} and Table~\ref{Tab:auc}).  On the other hand, as shown in this figure, DyNB performs close to DESeq2-GLM and worse than DESeq2-min, indicating its limitations to analyze temporal RNA-seq data from this GMNB generative model. This is due to the reason that the smooth assumption of DyNB may not always hold for the data generated by this gamma-Markov-chain based generative model. 

\subsubsection{Comparison based on DyNB generative model}

In the second simulation study, data is generated according to the DyNB model assumptions. More specifically, we draw the true mean values $\mu_k$, for $1000$ genes from a Gaussian process with the mean $m_k$ and the covariance matrix $Cov(t_i, t_j) = \theta_k \text{exp}(-\frac{1}{2\alpha_k}|t_i - t_j|)$, where $m_k$, $\theta_k$ and $\alpha_k$ are uniformly distributed in the intervals $[1000, 2000]$, $[100, 10000]$ and $[0.5, 1]$, respectively. We consider five time points at 0, 12, 24, 48 and 72 hours, similar to the real-world dataset \citep{aijo2014methods}. 10\% of genes are set to be truly differentially expressed across conditions by changing their mean values $m_k$ and covariance function parameters $\{ \theta_k, \alpha_k \}$ to $\{ b m_k, c \theta_k, \alpha_k \pm d \}$, where $b = 1.5$, $c = 10$, and $d = 0.25$ determine the significance of expected expression changes across conditions. Similar to the previous simulation setup, 4 replicates are generated for each time point in the corresponding condition.  %

\begin{figure}[t]
\begin{center}
\includegraphics[width=.85 \textwidth]{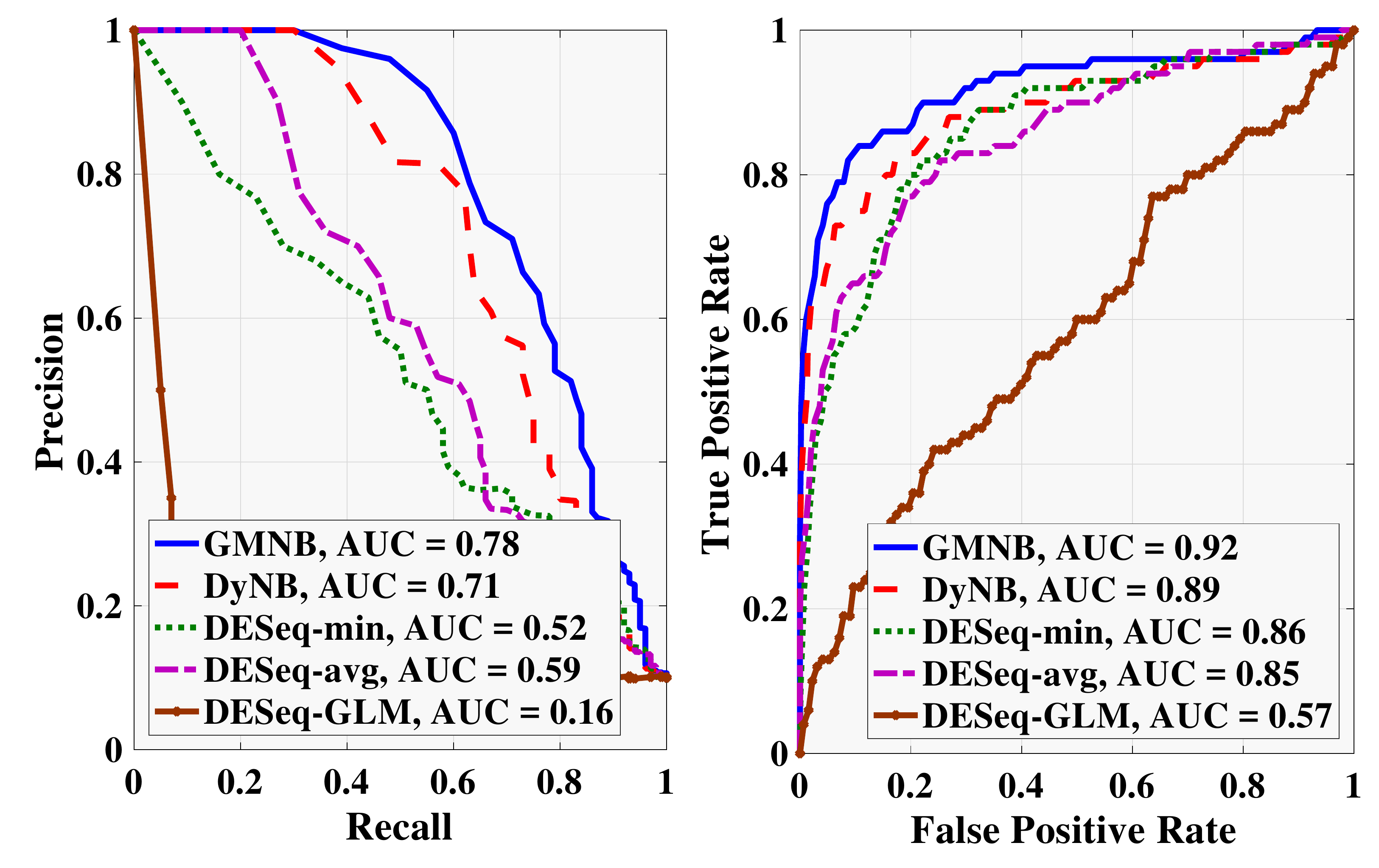}
\caption{{\bf Left column:} PR Curve, {\bf Right column:} ROC Curve. Performance comparison of different methods for differential gene expression over time based on the DyNB generative model. AUCs are given in the corresponding legends in the plots.}
\label{gp}
\end{center}
\end{figure}

Figure~\ref{gp} demonstrates the performances of different methods applied to the data generated according to the above procedure. GMNB still clearly outperforms the other methods based on both ROC and PR curves. The inferior performance of DyNB may be due to the small number of replicates for each time point, leading to poor estimates of both $\theta_k$ (heuristically estimated by the data dependent value $10 \times \mbox{stdev}(\mathbf{Y})$ based on the observed replicates $\mathbf{Y}=\{\mathbf{y_1, \dots, y_J}\}$ in \citet{aijo2014methods}) and $\mu_k$ (heuristically estimated by $\frac{\mbox{min}(\mathbf{Y}) + \mbox{max}(\mathbf{Y})}{2}$ in \citet{aijo2014methods}). On the contrary, the fully Bayesian nature of GMNB makes its performance robust to the number of replicates as well as potential noise at each time point. Both variates of static DESeq2 under-perform the dynamic approaches remarkably, as they neglect the correlation between samples across time points. In addition, DESeq2-GLM under-performs both GMNB and DyNB substantially, as it neglects the inherent dynamics of RNA-seq experiments specifically. %
To further demonstrate the potential influence of gene expression levels, reflected by the expected read counts, on the detection power of differentially expressed genes,
additional simulations are performed with the same parameters as above, except for the mean parameter $m_k$, for which three sampling uniform distributions are tested with the intervals $[1000, 2000]$, $[200, 1200]$, and $[50, 200]$, leading to three different datasets with different expected overall counts. We compare the differential expression analysis results for GMNB, DyNB, and DESeq2-min as they are top performing methods in this set of experiments for 20 randomly generated synthetic datasets for each setup. As shown in Figure~\ref{barplot}, according to both AUC-ROC and AUC-PR, GMNB consistently outperforms the other two methods no matter when we have low or high level counts. It is also noticeable that DyNB and DESeq2-min are more sensitive with variable performances across 20 randomly generated datasets. This indicates that GMNB better borrows signal strengths across time points compared to DyNB and DESeq2-min. %

\subsubsection{Comparison based on NB-AR(1) generative model}

In addition to synthetic data based on the GMNB and DyNB models, we evaluate these methods with the simulated data based on the NB-AR(1) model~\citep{oh2013time}. More precisely, the count for gene $k$ at time $t$ is distributed according to a NB distribution whose mean parameter satisfies $\text{log}(\mu_k^{(t)}) = \omega_k^{(t)} + \beta_k$. Here $\beta_k$ follows the uniform distribution in $[4.5, 5.5]$ to test the temporal differential expression performance with low read counts. The parameter $\omega_k^{(t)}$ is obtained through an auto-regressive process $\phi_k \omega_k^{(t-1)} + \epsilon^{(t)}$, where $\phi_k$ is randomly generated from the uniform distribution in $[0.1, 0.9]$, and $\epsilon^{(t)}$ is a standard zero-mean white noise process. Similar to the previous two simulation models, read counts are generated for $1000$ genes and 10\% of them selected to be differentially expressed by changing the parameter $\phi_k$ to $b \phi_k$ for the second condition, where
$$b =
  \begin{cases}
    3/2       & \quad \text{if } \phi_k \leq 0.5\\
    2/3  & \quad \text{if } \phi_k > 0.5
  \end{cases}
$$ determines the  significance of differential expression across conditions.

\begin{figure}[t]
\begin{center}
\includegraphics[width=.85 \textwidth]{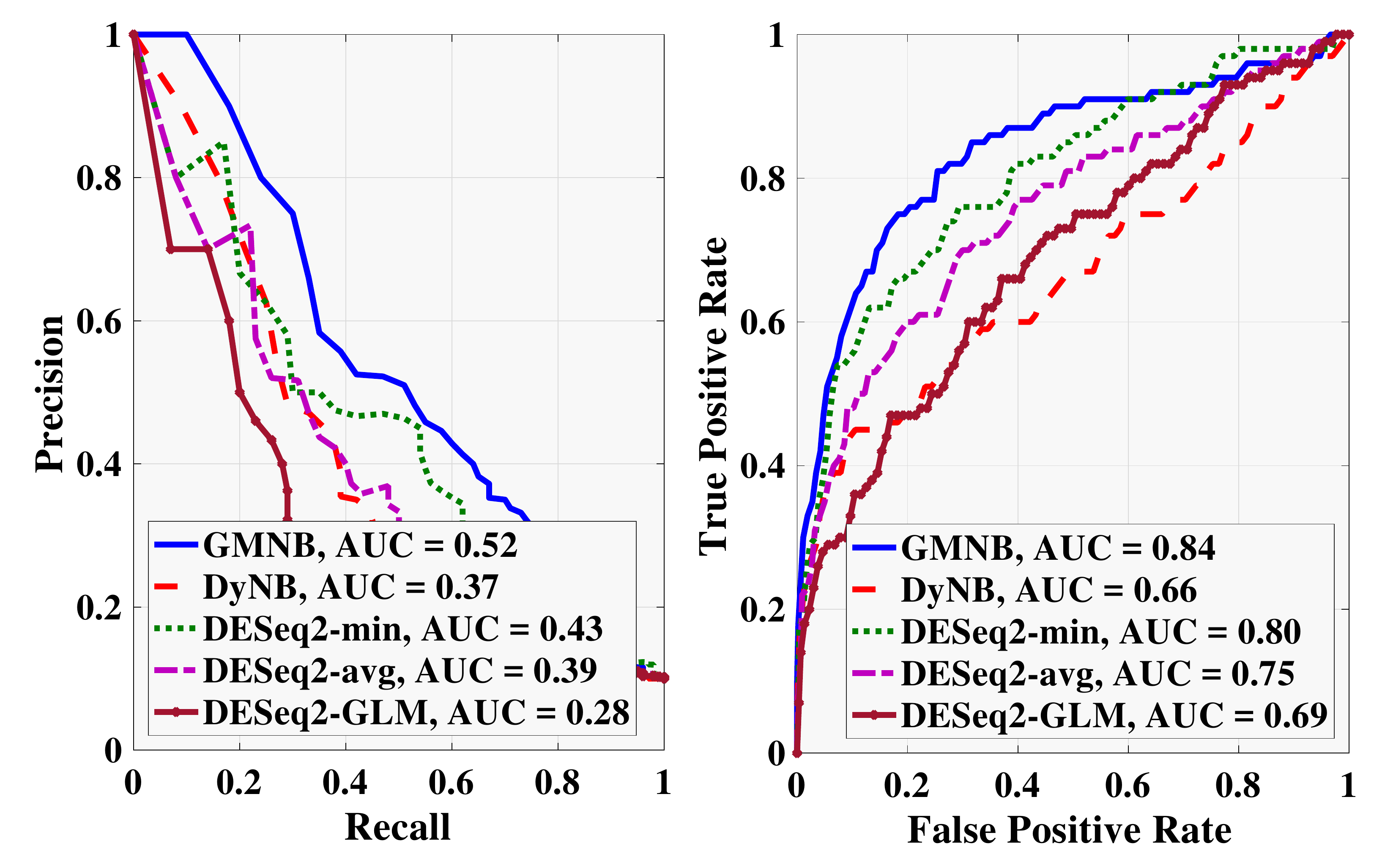}
\caption{{\bf Left column:} PR Curve, {\bf Right column:} ROC Curve. Performance comparison of different methods for differential gene expression over time based on the NB-AR(1) generative model. AUCs are given in the corresponding legends in the plots.}
\label{ar}
\end{center}
\end{figure}

\begin{figure}[h]
\begin{center}
\includegraphics[width=.85 \textwidth]{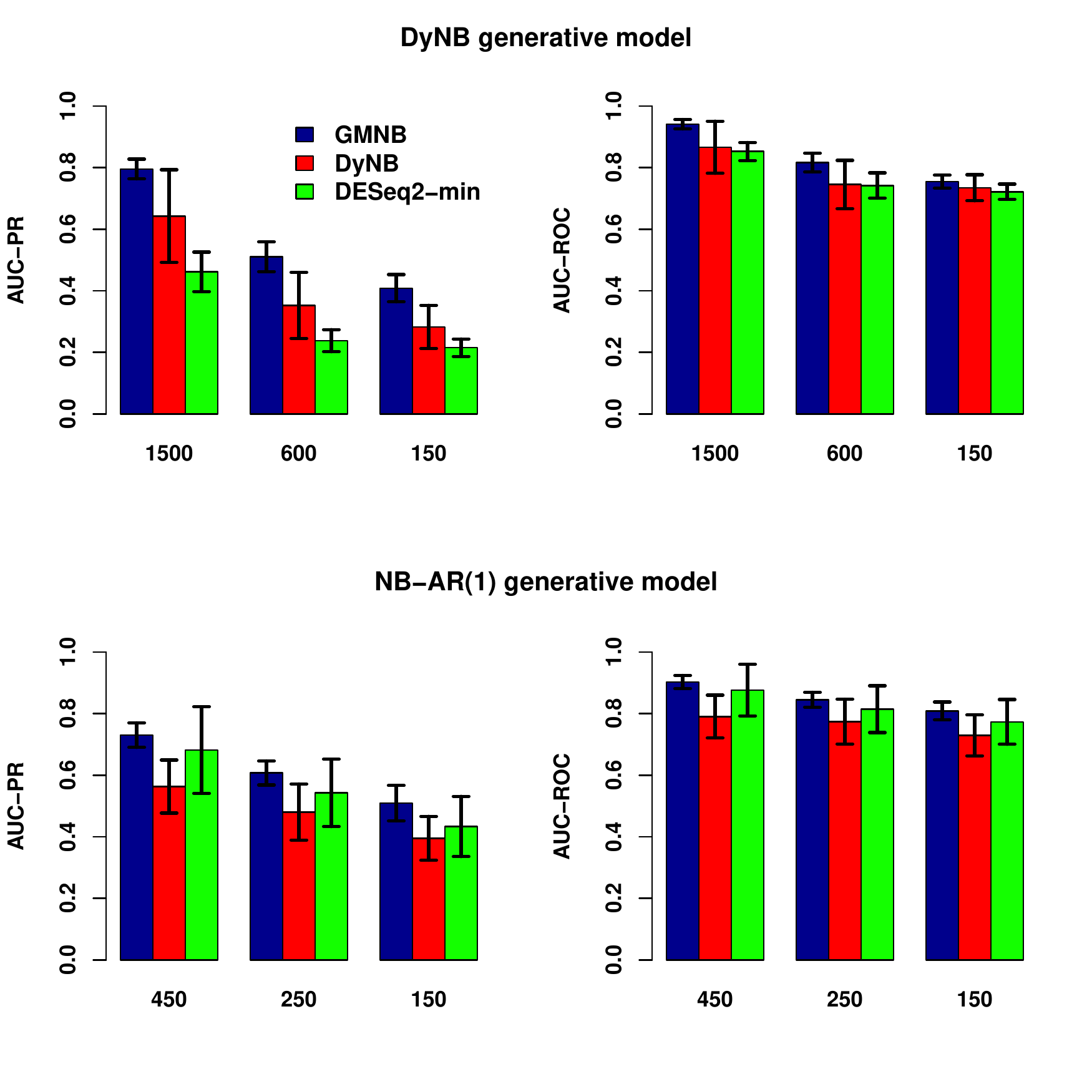}
\caption{AUC comparison of different methods for differential gene expression analysis over time in low counts.}
\label{barplot}
\end{center}
\end{figure}

Figure~\ref{ar} demonstrates the performances of different methods applied to the NB-AR(1) data. GMNB again outperforms DyNB and DESeq2 with a remarkable margin in both the ROC and PR curves. %
This is due to the state-space nature of the NB-AR(1) simulation setup, in which differential expression is defined through the model parameter $\phi_k$ that controls the temporal dependence of gene expression. However, the temporal correlation assumptions of the Gaussian process, %
different from this generative model, makes it less powerful to identify all differentially expressed genes. The results in Figure~\ref{ar} demonstrate the higher power of GMNB in detecting temporal differential expression patterns, especially with low expression levels (read counts are approximately 150 here). Similar to the DyNB generative model, we compare the performance of GMNB with DyNB and DESeq2-min (top performing methods) with different expected counts. In Figure~\ref{barplot}, additional simulations are performed with the same parameters as above, except the uniform distribution for the $\beta_k$ with three varying  intervals: $[4.5, 5.5]$, $[4.5, 6.5]$, and $[5.5, 6.5]$, leading to corresponding expected counts from 150 to 450. Figure~\ref{barplot} shows again that for low counts, GMNB outperforms both DyNB and DESeq2, especially in AUC-PR. Note that, with increasing expression levels, DESeq2-min's performance improves because of high signal strengths at individual time points. 
As shown by the ROC and PR curves in both the GMNB and AR generative models, DESeq2-min outperforms DyNB. This indicates that the temporal correlation assumptions in DyNB may not fully capture the dynamic changes in these two state-space generative models, which can have abrupt non-smooth changes. In addition, the heuristic estimation of model parameters adopted in DyNB \citep{aijo2014methods} when the number of replicates is low can be the other reason for the degraded performance.

In summary, on synthetic RNA-seq count data from different generative models, comparison of both the ROC and PR curves shows that GMNB outperforms both the recently proposed temporal (DyNB) and static differential analysis methods that aggregate differential statistics in heuristic ways (DESeq2 with different setups). Table~\ref{Tab:auc} summarizes the average AUCs and their standard deviation values of both ROC and PR curves for 20 randomly generated synthetic datasets by the top three performing methods (GMNB, DyNB, and DESeq2-min). 
GMNB improves the performances of DyNB and DESeq2-min, in terms of AUC-PR, at least by $23\%$ and $17\%$, respectively. In the best case scenario, GMNB improves the AUC-PR performances of DyNB and DESeq2-min up to $48\%$ and $71\%$, respectively. In terms of AUC-ROC, GMNB improves the best case performances of DyNB and DESeq2-min by $12\%$ and $10\%$, respectively. 
Even with the data from the DyNB generative model, the fully Bayesian method GMNB outperforms DyNB, which estimates some of its model parameters in a heuristic manner \citep{aijo2014methods}. In addition, GMNB achieves robust performance in both state-space (GMNB and NB-AR(1)) and functional (DyNB) generative models. We demonstrate the superior power of GMNB in low count situations by collective information across time points. For these three different types of synthetic data, as shown in Figures~\ref{gmnb} - \ref{barplot}, and Table~\ref{Tab:auc}, measured by
both AUC-ROC and AUC-PR, it is interesting to notice that DyNB works better than DESeq2-min only when the synthetic data are generated based on its model assumption.

\begin{table}[!t]
\centering
\caption{Comparison of AUCs based on 20 runs for each method.\label{Tab:auc}}
{\begin{tabular}{@{}l|l|lll@{}}
\toprule {\bf AUC} & {\bf Generative Model} & {\bf GMNB } & {\bf DyNB} & {\bf DESeq2-min} \\\hline \hline
{} & {\bf GMNB} & {\bf 0.84 $\pm$ 0.02} & {0.75 $\pm$ 0.05} & {0.80 $\pm$ 0.07}\\
{\bf ROC} &{\bf DyNB} & {\bf 0.94 $\pm$ 0.01} & {0.86 $\pm$ 0.08} & {0.85 $\pm$ 0.03}\\
{} & {\bf NB-AR(1)} & {\bf 0.81 $\pm$ 0.03} & {0.73 $\pm$ 0.07} & {0.77 $\pm$ 0.07}\\\hline
{} & {\bf GMNB} & {\bf 0.61 $\pm$ 0.04} & {0.41 $\pm$ 0.08} & {0.52 $\pm$ 0.06}\\
{\bf PR} & {\bf DyNB} & {\bf 0.79 $\pm$ 0.03} & {0.64 $\pm$ 0.20} & {0.46 $\pm$ 0.06}\\
{} & {\bf NB-AR(1)} & {\bf 0.51 $\pm$ 0.06} & {0.39 $\pm$ 0.07} & {0.43 $\pm$ 0.10}\\\hline
\end{tabular}}{}
\end{table}

\subsection{Human Th17 cell induction}

To further illustrate how GMNB may help identify differentially expressed genes from temporal RNA-seq data for biologically significant results, we provide such a case study consisting of $57$ human samples during the priming of T helper 17 (Th17) cell differentiation \citep{tuomela2012identification}. The main goal of designing this case study is to gain insights into the differentiation process by unraveling dependency between different genetic factors in various pathways, which may serve as potential biomarkers of immunological diseases for therapeutic intervention design. In this dataset \citep{tuomela2016comparative}, 
at $0, 0.5, 1, 2, 4, 6, 12, 24, 48$, and $72$ hours of Th17 polarized cells and control Th0 cells, three biological replicates were collected for transcript profiling by RNA-seq. The data were downloaded from Gene Omnibus with the accession number GSE52260~\citep{tuomela2016comparative,chan2016subpopulation}. %

When checking the 10 %
most
differentially expressed genes based on their BFs by GMNB, all of them have been reported to be differentially expressed in other studies investigating Th17 cell differentiation. Among them, the top differentially expressed gene is thrombospondin-1 (TSP1), whose encoded protein participates in the differentiation of Th17 cells by activating transforming growth factor beta (TGF-$\beta$) and enhancing the inflammatory response in experimental autoimmune encephalomyelitis (EAE) \citep{yang2009deficiency}. 
The second gene in the list is Lymphotoxin $\alpha$ (LTA), a member of the tumor necrosis factor (TNF) superfamily
that is both secreted and expressed on the cell surface of activated Th17 cells \citep{chiang2009targeted}.
The third gene, COL6A3, contributes to adipose tissue inflammation \citep{pasarica2009adipose}
and responds quickly to Th17 cell polarizing stimulation \citep{tripathi2017genome}.
The gene Cathepsin L (CTSL1) is ranked as the fourth in the list and is linked to the regulation of immune responses at the level of MHC complex maturation and Ag presentation influencing differentiation of CD4+ cells and autoimmune reactions \citep{reiser2010specialized}. The fifth gene, FURIN, has been reported as a T cell activation gene that regulates the T helper cell balance of the immune system \citep{pesu2008t}. The sixth gene lamin A (LMNA) has been identified as one of the immune response regulators \citep{gonzalez2014nuclear}. The seventh gene, Filamin A (FLNA), is required for T cell activation \citep{hayashi2006filamin}. The eighth gene, SBNO2, has been reported to influence Th17 cell differentiation \citep{tripathi2017genome}. \citet{zhao2014comparison} observed significant changes in the expression of the ninth gene ACTB in activated T cells.
Finally, the tenth gene Notch1 is activated in both mouse and human in vitro-polarized Th17 cells and also  in Th17 polarized cells as compared with Th0 control cells \citep{keerthivasan2011notch}.

\begin{figure}[htp]
 
\begin{subfigure}{0.5\textwidth}
\includegraphics[width=\textwidth]{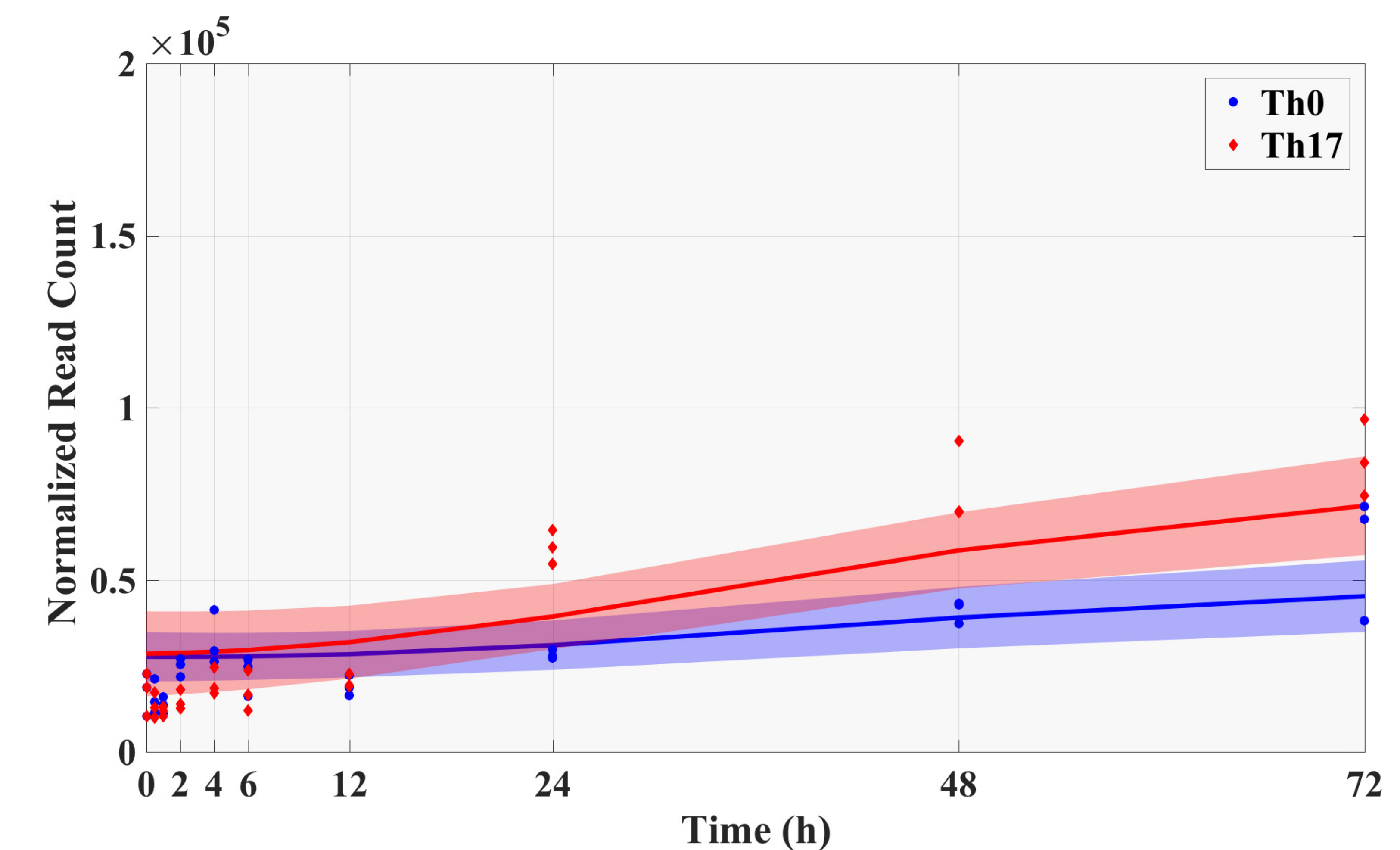} 
\caption{}
\label{g9_norm}
\end{subfigure}
\begin{subfigure}{0.5\textwidth}
\includegraphics[width=\textwidth]{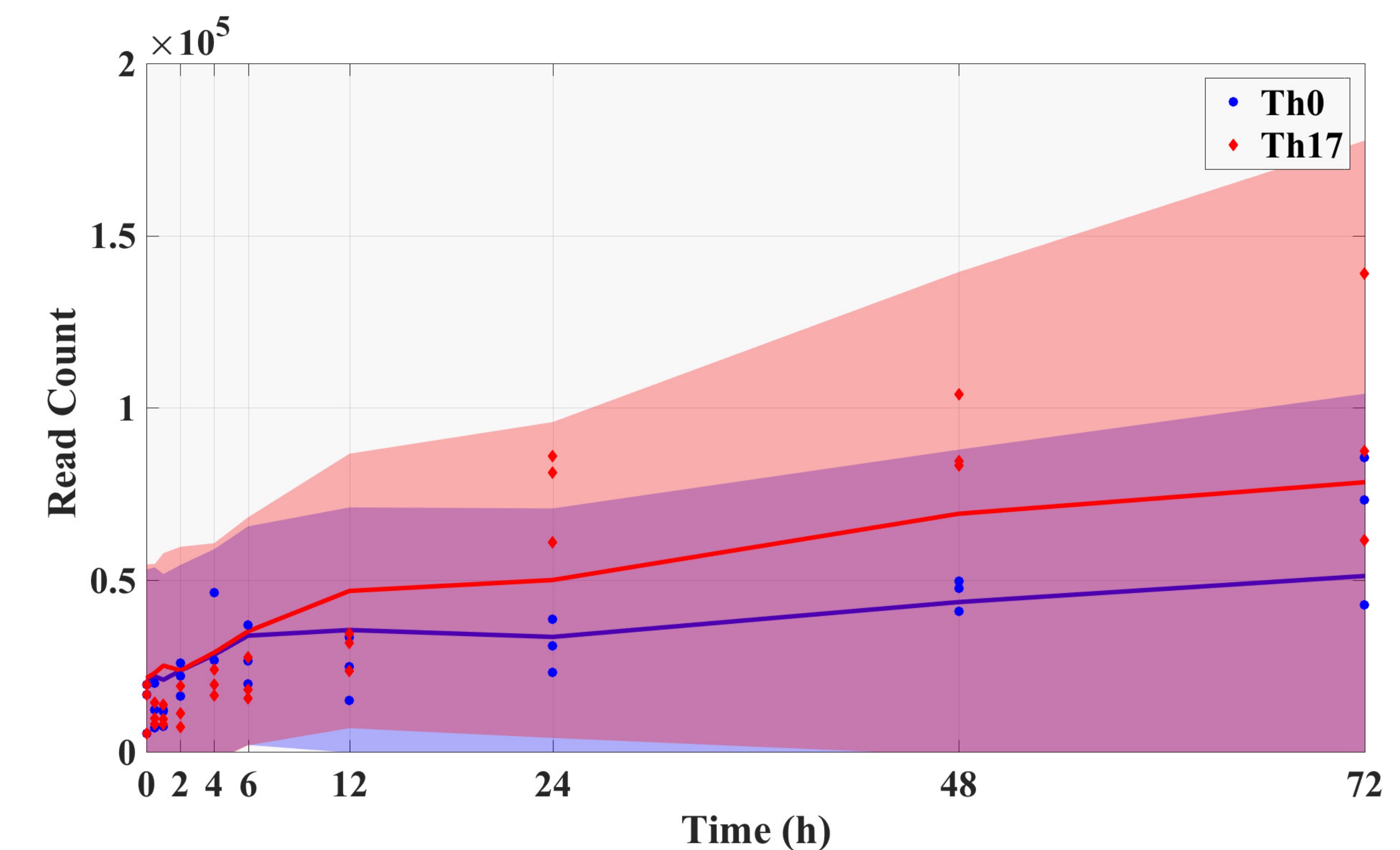}
\caption{}
\label{g9_raw}
\end{subfigure}
\begin{subfigure}{0.5\textwidth}
\includegraphics[width=\textwidth]{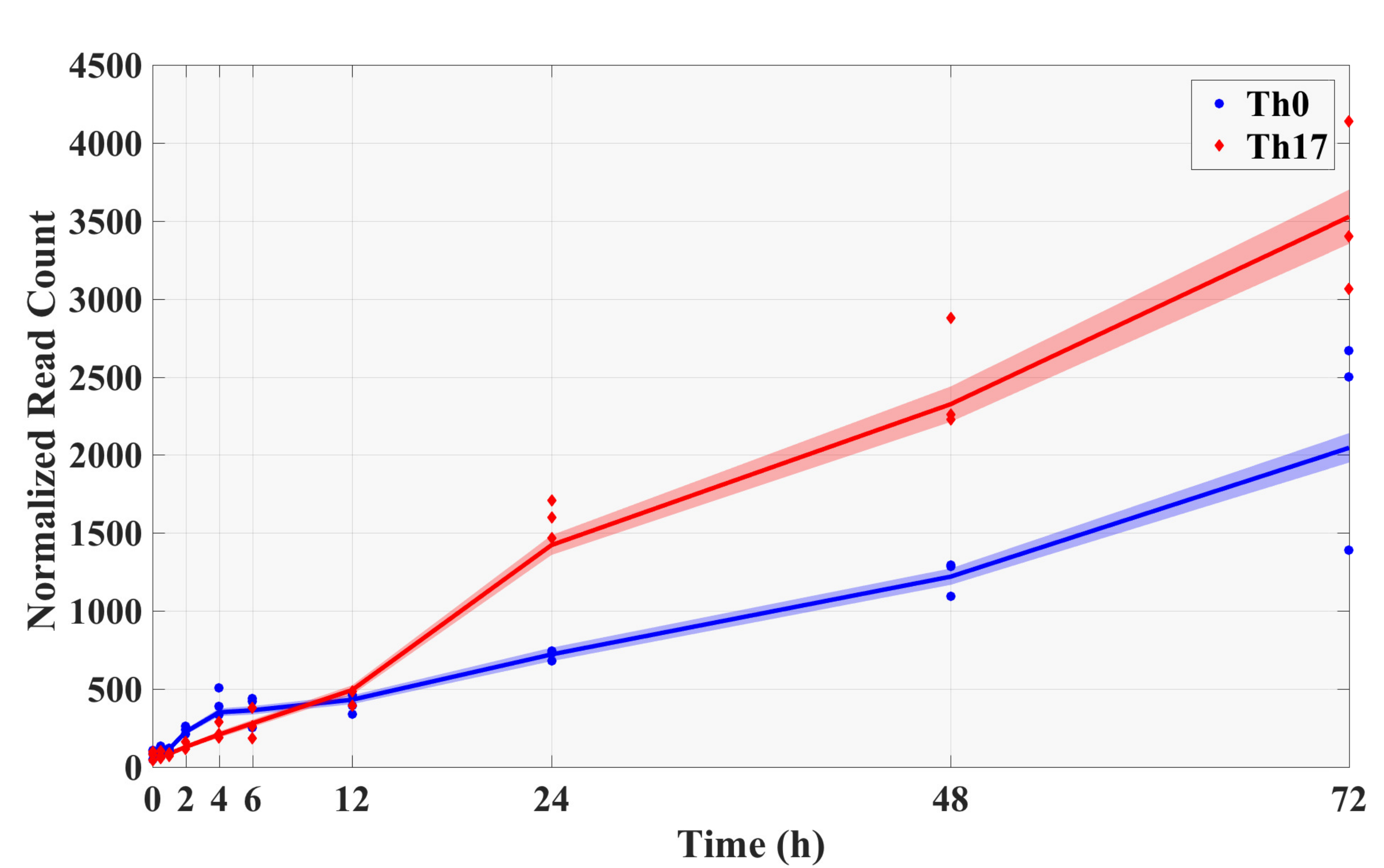} 
\caption{}
\label{g9_gmnb_norm}
\end{subfigure}
\begin{subfigure}{0.5\textwidth}
\includegraphics[width=\textwidth]{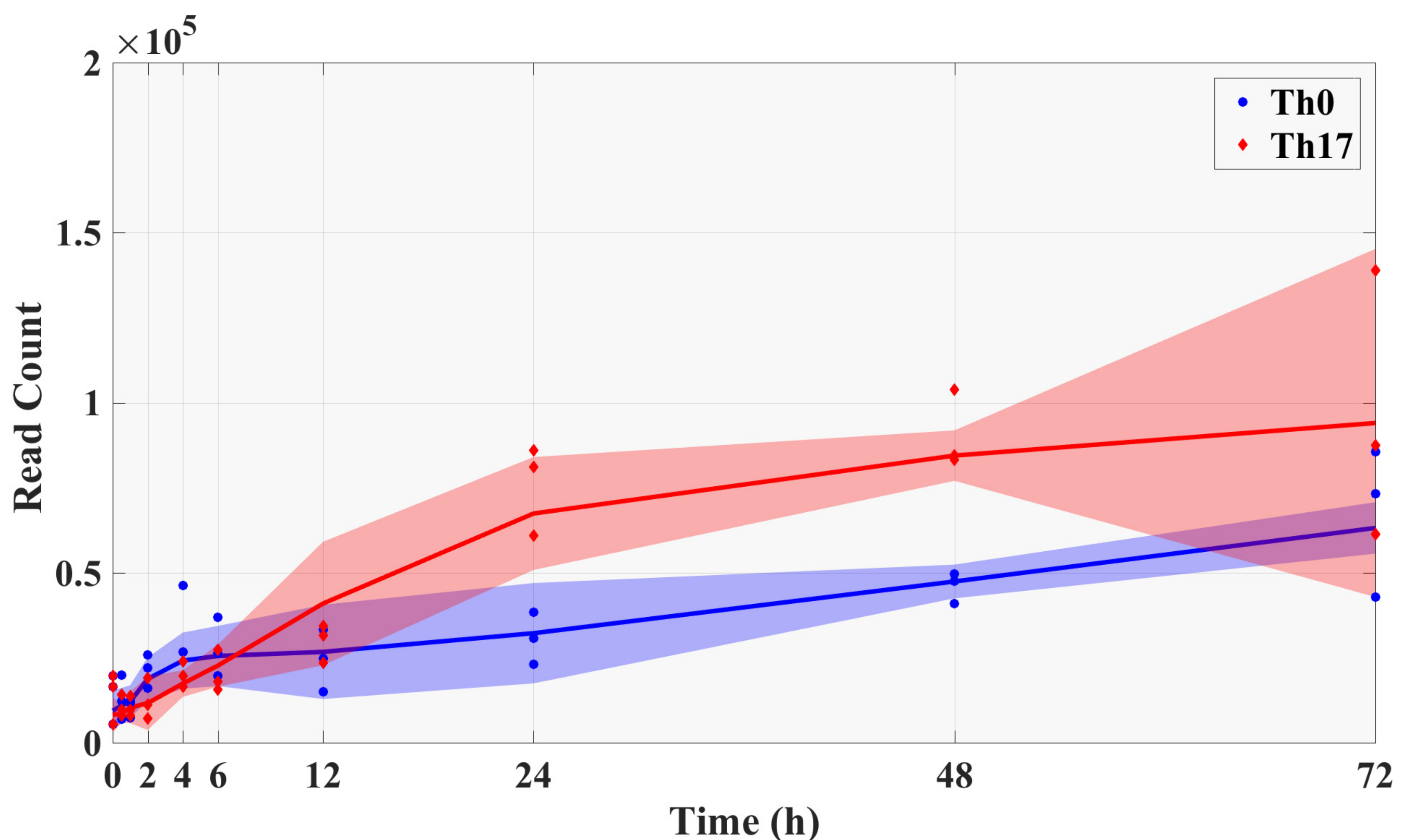}
\caption{}
\label{g9_gmnb_raw}
\end{subfigure}
 
\caption{\textbf{Differentially expressed gene FLNA detected by GMNB but not by DyNB.} (a) The normalized gene expression profile of {\bf FLNA} over time estimated by DyNB model. The normalization of read counts on the y-axis are obtained by using the normalization method of DESeq. The solid blue and red curves are the posterior means under Th0 and Th17 lineages, respectively, with corresponding $99\%$ CIs (shaded areas). (b) The gene expression profile of \textbf{FLNA} over time estimated by DyNB. The read counts on the y-axis are observed read count. The solid blue and red curves are the means of the generated samples based on the inferred parameters by DyNB under Th0 and Th17 lineages, respectively, with corresponding $99\%$ CIs (shaded areas around means). (c) The normalized gene expression profile of {\bf FLNA} over time estimated by the proposed GMNB model. The normalization of read counts on the y-axis are obtained by dividing the observed counts by the parameter ${p_j^{(t)}}/({1-p_j^{(t)})}$ representing the sequencing depth in the model. The solid blue and red curves are posterior means of $r_k$ under Th0 and Th17 lineages, respectively, with corresponding $99\%$ CIs (shaded areas). (d) The gene expression profile of \textbf{FLNA} over time estimated by GMNB. The read counts on the y-axis are observed read count. The solid blue and red curves are the means of the generated samples based on the inferred parameters by GMNB under Th0 and Th17 lineages, respectively, with corresponding $99\%$ CIs (shaded areas around means).}
\label{g9}
\end{figure}

We then investigate how the results of DyNB differ from those of GMNB. The majority of the above genes are indeed ranked relatively high by DyNB as differentially expressed, except two genes: FLNA and ACTB. For these two genes, their expression levels change abruptly after 12 hours of T17 differentiation. These two genes demonstrate that the DyNB method may fail to detect temporal differential expression when the temporal gene expression trends are not smooth. As an instance, Figure~\ref{g9} illustrates that DyNB is not able to capture the temporal expression changes of gene FLNA accurately. More precisely, Figure~\ref{g9}(a) shows the posterior means of expected gene expression $\mu_k$ based on DyNB and their corresponding confidence intervals, where circles and diamonds represent the normalized counts from Th0 and Th17 lineages, respectively. To further assess the power of the models in reproducing the observed gene counts, for each model, we generate 1000 gene counts per sample and time point based on the inferred parameters, and then calculate the 99\% confidence interval using these synthetically generated counts. Figure~\ref{g9}(b) demonstrates the means and confidence intervals of the counts generated via this procedure for DyNB, where the circles and diamonds represent the observed raw counts from Th0 and Th17 lineages, respectively. Similar to plots in Figures~\ref{g9}(a) and ~\ref{g9}(b), we perform the same examinations on expression pattern of FLNA by the GMNB model. To demonstrate the expression levels of the $k$th gene between two groups, the DyNB compares the posterior NB mean parameters $\mu_k$, whereas the GMNB compares the posterior NB shape parameters $r_k$.
One may consider that the expression level of gene $k$ is assumed to roughly follow a function of the shape parameter $r_k$ in the GMNB, but the observed counts should be demonstrated in a same scale as the shape parameter. The difference between the posterior shape parameters $r_k$ explains the differences between the means, since if $n_{kj}^{(t)} \sim \mbox{NB}(r_k^{(t)}, p_j^{(t)} )$, then $\mathop{\mathbb{E}} [n_{kj}^{(t)}] = r_k^{(t)}p_j^{(t)}/(1 - p_j^{(t)})$.
Therefore, Figure~\ref{g9}(c) shows the posterior means of $r_k$ based on GMNB and their corresponding confidence intervals, where the circles and diamonds are obtained by dividing the observed counts by the parameter $p_j^{(t)}/(1-p_j^{(t)})$ representing the sequencing depth in the proposed model. Additionally, Figure~\ref{g9}(d) demonstrates the means and confidence intervals for synthetically generated gene counts based on the inferred parameters of GMNB, where the read counts on the y-axis are observed read counts. Not only GMNB improves the model fitting over 24h to 72h, but also it has more robust estimation of expression patterns for the starting time points with lower counts (Figures~\ref{g9}(c) and~\ref{g9}(d)). The calculated BFs for the gene FLNA are 2.3461 and $1.60 \times 10^{308}$ by DyNB and GMNB, respectively.
GMNB also identifies ACTB as a gene with significant differential temporal expression (BF $>$ 10) but DyNB again fails to capture the abrupt expression changes and thereby associates low BF (supplement materials). The corresponding temporal expression plots are depicted in Figure~S1 of the supplement materials.%

\begin{figure}[htp]
 
\begin{subfigure}{0.5\textwidth}
\includegraphics[width=\textwidth]{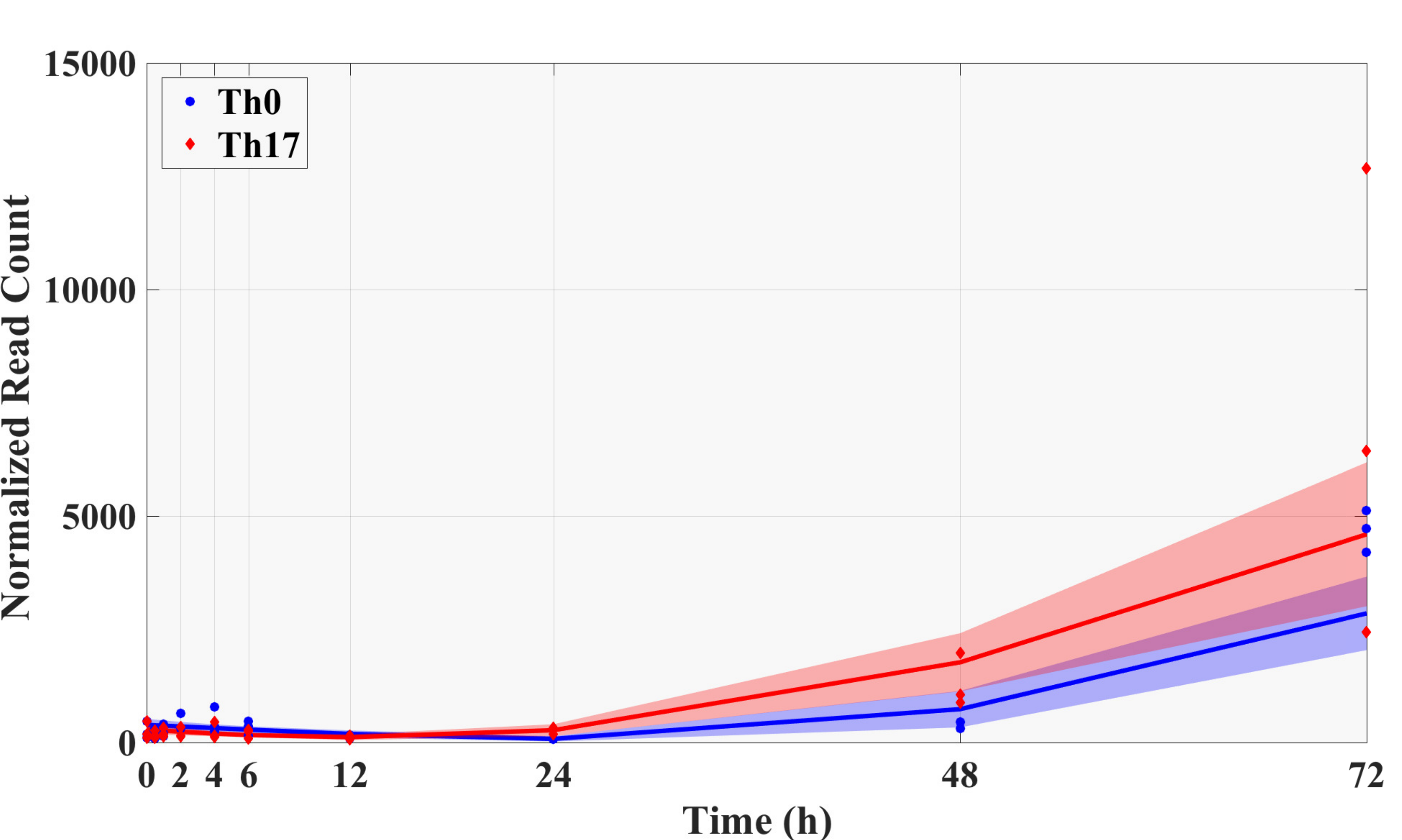} 
\caption{}
\label{lgals_norm}
\end{subfigure}
\begin{subfigure}{0.5\textwidth}
\includegraphics[width=\textwidth]{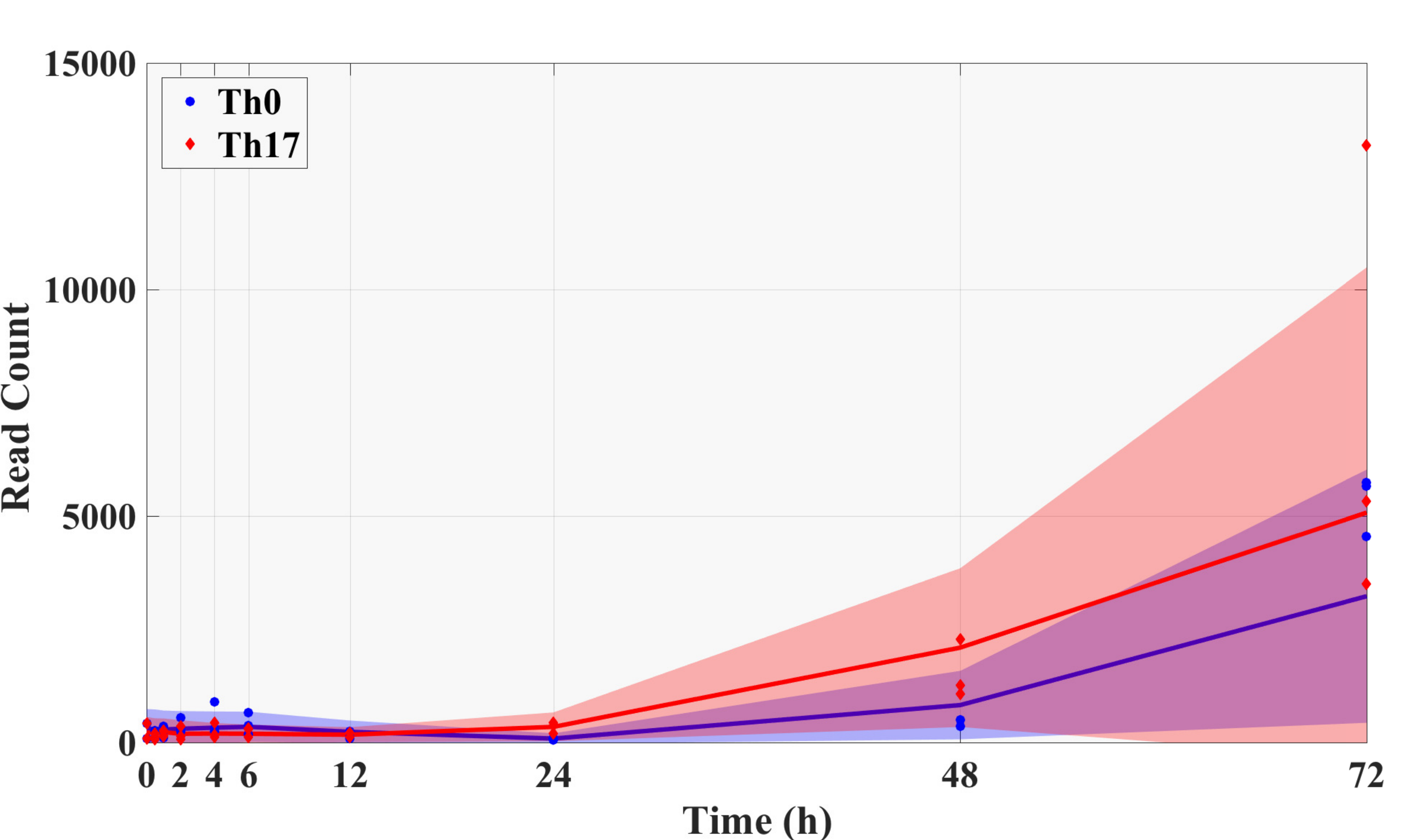}
\caption{}
\label{lgals_raw}
\end{subfigure}
\begin{subfigure}{0.5\textwidth}
\includegraphics[width=\textwidth]{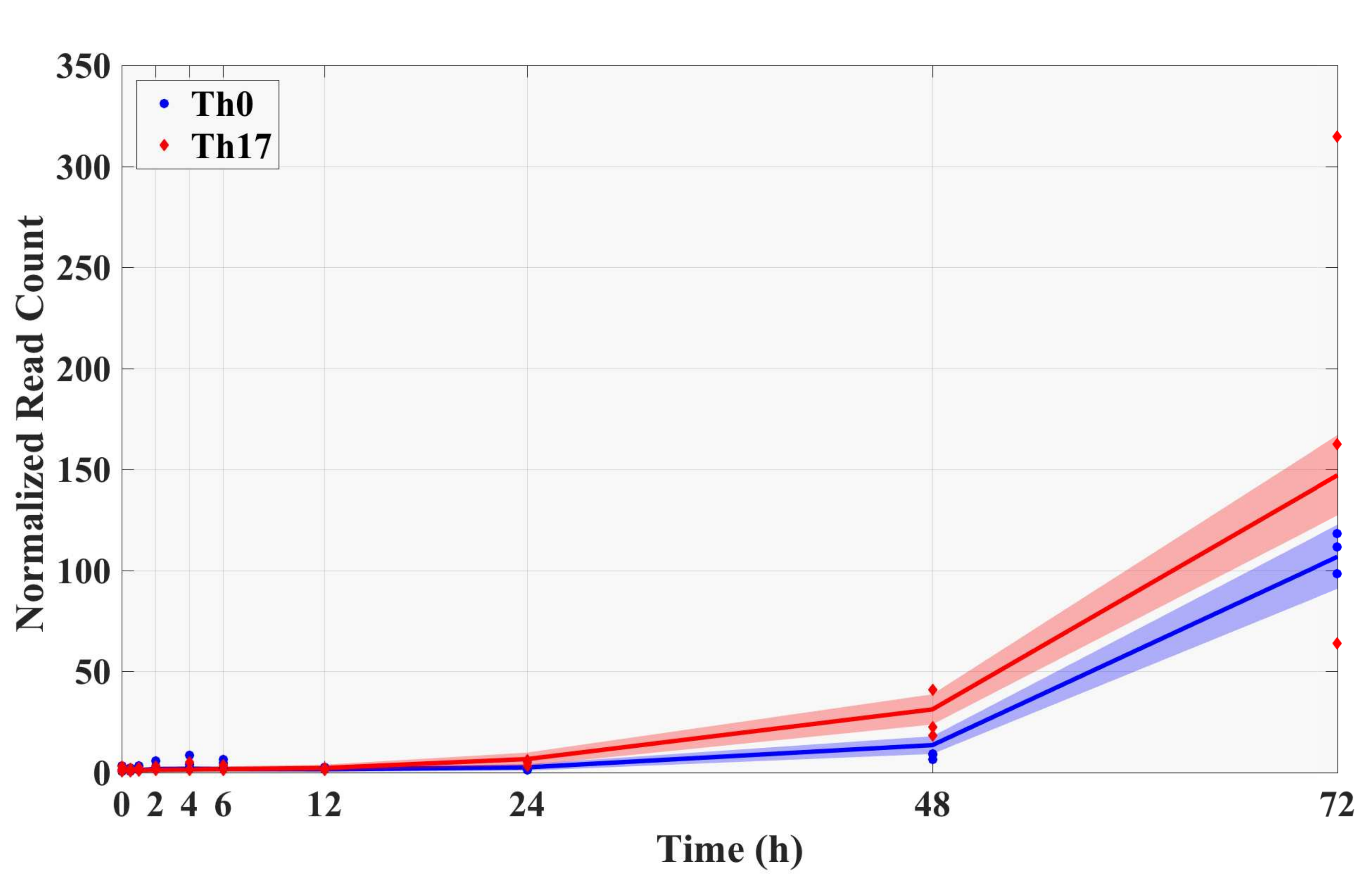} 
\caption{}
\label{lgals_gmnb_norm}
\end{subfigure}
\begin{subfigure}{0.5\textwidth}
\includegraphics[width=\textwidth]{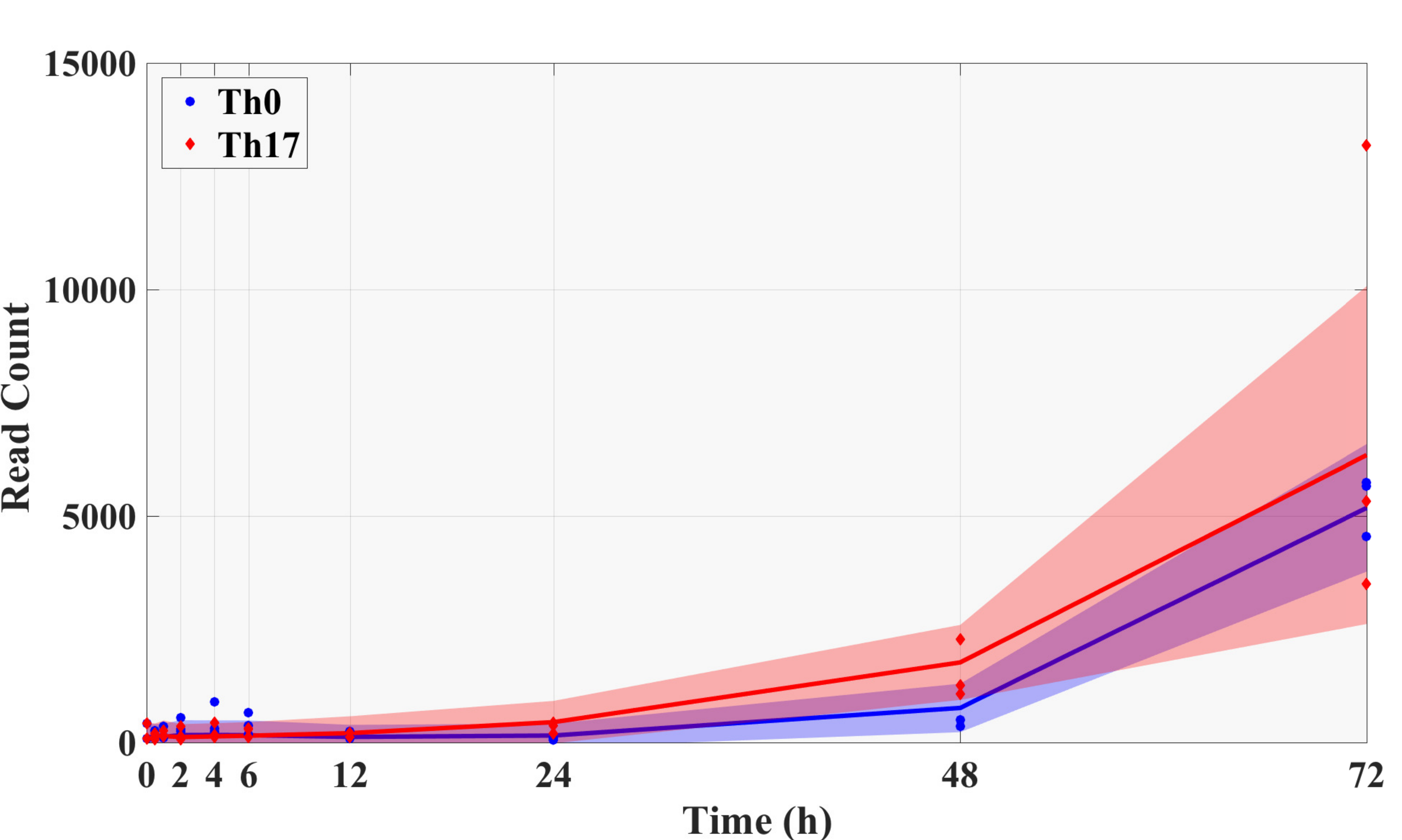}
\caption{}
\label{lgals_gmnb_raw}
\end{subfigure}
 
\caption{\textbf{Example of genes detected as differentially expressed by DyNB but not by GMNB: \textbf{LGALS1}.} (a) The normalized gene expression profile of {\bf LGALS1} over time estimated by DyNB model. The normalization of read counts on the y-axis are obtained by using the normalization method of DESeq. The solid blue and red curves are the posterior means under Th0 and Th17 lineages, respectively, with corresponding $99\%$ CIs (shaded areas). (b) The gene expression profile of \textbf{LGALS1} over time estimated by DyNB. The read counts on the y-axis are observed read count. The solid blue and red curves are the means of the generated samples based on the inferred parameters by DyNB under Th0 and Th17 lineages, respectively, with corresponding $99\%$ CIs (shaded areas around means). (c) The normalized gene expression profile of {\bf LGALS1} over time estimated by the proposed GMNB model. The normalization of read counts on the y-axis are obtained by dividing the observed counts by the parameter ${p_j^{(t)}}/({1-p_j^{(t)})}$ representing the sequencing depth in the model. The solid blue and red curves are posterior means of $r_k$ under Th0 and Th17 lineages, respectively, with corresponding $99\%$ CIs (shaded areas). (d) The gene expression profile of \textbf{LGALS1} over time estimated by GMNB. The read counts on the y-axis are observed read count. The solid blue and red curves are the means of the generated samples based on the inferred parameters by GMNB under Th0 and Th17 lineages, respectively, with corresponding $99\%$ CIs (shaded areas around means).}
\label{lgals}
\end{figure}

On the other hand, LGALS1, SEPT5, BATF3, COL1A2, and ENO2 are five genes out of 90 differentially expressed genes detected by DyNB with BFs $2.59 \times 10^{7}$, $472.34$, $2.90 \times 10^{4}$, $404.34$, and $398.43$, whereas they are associated with BFs lower than 10 by GMNB. %
Figure~\ref{lgals} illustrates the expression profile of the gene LGALS1 inferred by DyNB and GMNB, indicating that DyNB is not able to filter out those low count genes for which the replicated Th0 and Th17 lineages are seemingly similar and leads to this potential false positive. Figure~\ref{lgals}(a) shows the posterior means of expected gene expression $\mu_k$ based on DyNB under Th0 and Th17 lineages with their corresponding confidence intervals. Figure~\ref{lgals}(b) shows the means and confidence intervals for 1000 generated samples based on the inferred parameters of DyNB model. While the normalized counts are plotted in Figure~\ref{lgals}(a), the circles and diamonds mark Th0 and Th17 lineages, respectively, for the observed raw counts in Figure~\ref{lgals}(b). On the contrary, GMNB considers this gene not significantly differentially expressed with similar inferred temporal expression profiles across conditions, as demonstrated in Figures~\ref{lgals}(c) and~\ref{lgals}(d). This may be explained by the fact that GMNB employs a fully generative model of gene expressions, including the sequencing depth, while DyNB uses a deterministic ad-hoc procedure to normalize gene counts, and thus neglecting the uncertainty over the sequencing depth when computing the BF, leading to potential false positives.
Figures~\ref{lgals}(c) and~\ref{lgals}(d) demonstrate the posterior means of $r_k$ based on GMNB and the means of synthetically generated samples based on the inferred parameters of the proposed model, respectively. %
Similar to the plots for LGALS1, Figures~S8,~S9,~S10, and~S11 in the supplement materials show the similar trends for the genes SEPT5, BATF3, COL1A2, and ENO2 based on the results by DyNB and GMNB.

\begin{figure}
\begin{center}
\includegraphics[width=.5 \textwidth]{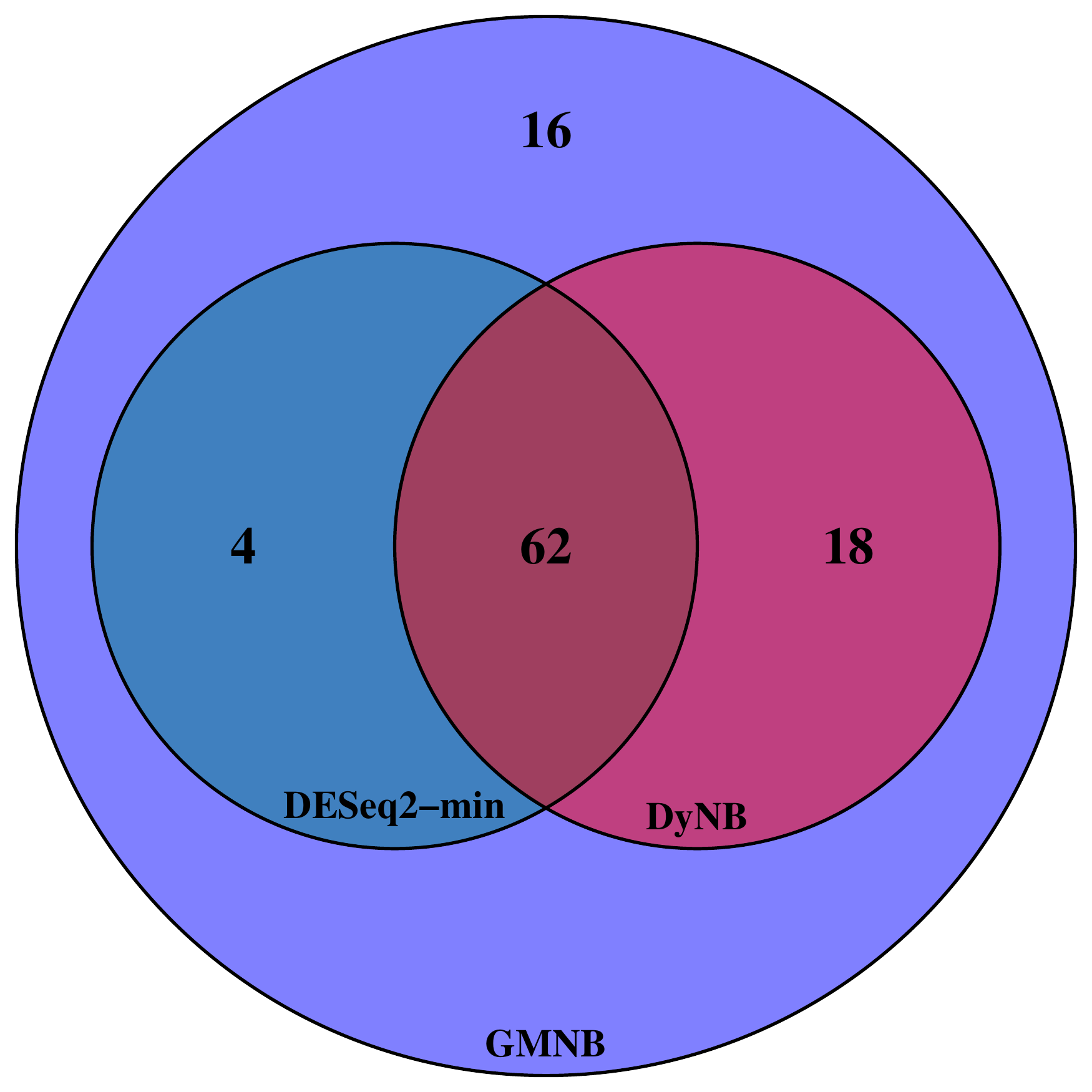}
\caption{A Venn diagram representing the overlaps of the top 100 differentially expressed genes detected by GMNB with DyNB and DESeq2-min.}
\label{venn}
\end{center}
\end{figure}

In order to further demonstrate the advantages of GMNB, the overlap of three approaches (GMNB, DyNB and DESeq2-min), for 100 top differentially expressed genes identified by GMNB, is depicted as a Venn diagram in Figure~\ref{venn}. A gene is differentially expressed based on DESeq2-min if the corresponding $\text{p-value} < 0.05$ at any time point. Out of top 100 differentially expressed genes identified by GMNB ($\mbox{log(BF)} > 100$), 16 genes are identified only by GMNB. The temporal expression plots for six of them, i.e. the genes EGR1, NR4A1, MYC, PKM2, EGR2, and IL6ST, are depicted in %
Figures~S2,~S3,~S4,~S5,~S6, and~S7, indicating the differential dynamic patterns identified by GMNB.
Among these genes, EGR1 is a transcription factor known to inhibit the expression of GFI1, a negative regulator of Th17 differentiation, by directly binding to its promoter and its expression is detected only in the early phase of Th17 differentiation \citep{kurebayashi2012pi3k}. The gene NR4A1 plays critical roles in T cell apoptosis during the thymocyte development \citep{doi2008orphan}. Not only this gene is a proapoptotic transcription factor, but also it is reported as a survival factor and activator of metabolic pathways. Both facets show the NR4A1's role in T-cell differentiation as a balancing molecule in the fate determination \citep{fassett2012nuclear}.
The gene MYC has been reported as one of the key transcript factors for Th17 differentiation \citep{yosef2013dynamic,sawcer2011genetic,gnanaprakasam2017myc}. PKM2 is induced and interacts with and promotes the function of HIF1$\alpha$ that is critical to drive Th17 differentiation \citep{corcoran2016hif1alpha}. EGR2 has been identified as an important transcription factor in the development and function of Th17 cells \citep{zhang2015roles,zhu2008early}. IL6ST is known as a signature transcript of Th17 cells \citep{ghoreschi2010generation}.
This again illustrates the benefits of GMNB on better modeling temporal dynamic changes to detect biologically meaningful genes who show significant difference in temporal changes but do not show significant differential expression when studying them at individual time points.

\subsubsection{Gene Ontology (GO) analysis}
To further demonstrate the biological relevance of the detected genes by GMNB, 
GO analysis of top 100 differentially expressed genes ($\mbox{log(BF)} > 100$) has been performed using Fisher's exact test. Enriched GO terms (Table S1 in the supplement materials) by these genes agree
with the current biological understanding of the Th17 differentiation process.
The most significantly enriched GO terms are related to the organ development ($\text{p-value} <  2 \times 10^{-23}$), immune system process ($\text{p-value} < 6 \times 10^{-21}$), immune response ($\text{p-value} < 1 \times 10^{-19}$), response to stimulus ($\text{p-value} < 3 \time 10^{-19}$), cell differentiation ($\text{p-value} < 3 \times 10^{-18}$), and defense response ($\text{p-value} < 2 \times 10^{-16}$).
In particular, $38\%$ and $74\%$ of these 100 genes are annotated to immune response and response to stimulus, respectively, supported by the central role of Th17 cells in the pathogenesis of autoimmune and inflammatory diseases \citep{waite2011th17}. %

\subsection{RNA-seq data in \citet{aijo2014methods}: Human-activated T- and Th17 cells}
\label{sec:dynbds}

We further analyze the second temporal RNA-seq dataset, for which DyNB was implemented for studying Th17 cell lineage \citep{aijo2014methods}.
In this dataset,  CD4+ T cells were activated and polarized as described in \citet{tuomela2012identification} and RNA-seq data were collected at 0, 12, 24, 48 and 72 hours of both the activation (Th0) and differentiation (Th17).
At each time point, there are 3 biological replicates for both cell lineages.
The original paper \citep{aijo2014methods} performed DyNB to quantify Th17-specific gene expression dynamics. 

The authors in~\citet{aijo2014methods} first normalized the RNA-seq counts by the DESeq pipeline \citep{anders2010differential}. Then, DyNB was applied to the normalized expression values to identify differentially expressed genes between the Th0 and Th17 lineages. Genes were considered differentially expressed if (i) $\mbox{BF} > 10$, and (ii) $\mbox{fold-change} > 2$ for at least one time point. Out of 698 differentially expressed genes identified by DyNB, three genes were investigated and discussed in~\citet{aijo2014methods} with the qRT-PCR validation: $IL17A$, $IL17F$, and $RORC$.

We apply GMNB to analyze the same Th17 cell lineage dataset to identify differentially expressed genes. To compare the ranked lists of genes by GMNB and by DyNB respectively, Table 1 gives the ranks as well as the computed BF values by GMNB and DyNB for these reported genes in \citet{aijo2014methods}. These qRT-PCR validated genes are in fact ranked higher by GMNB, indicating more promising potential for marker gene identification.

\begin{table}[!t]
    \centering
	\caption{Comparison of BF ranks for the reported genes by DyNB\label{Tab:01}}
{\begin{tabular}{@{}l||ll@{}}
\toprule {\bf Genes} & {\bf DyNB } & {\bf GMNB}  \\\hline \hline
{\bf RORC} & 37 (BF = $2.26 \times 10^{93}$) & {\bf 26} (BF = $2.98 \times 10^{48}$) \\
{\bf IL17F} & 352 (BF = $1.74 \times 10^{15}$) & {\bf 175} (BF = $2.53 \times 10^{9}$)\\
{\bf IL17A} & 755 (BF = $6.96 \times 10^{8}$) & {\bf 345} (BF = $3.90 \times 10^{4}$) \\\hline
\end{tabular}}{}
\end{table}

\section{Conclusions}

GMNB offers a comprehensive and fully Bayesian solution to study temporal RNA-seq data. The most notable advantage is the capacity to capture a broad range of gene expression patterns over time by the integration of a gamma Markov chain into a negative binomial distribution model. This allows GMNB to offer consistent performance over different generative models and makes it be robust for studies with different numbers of replicates by borrowing the statistical strength across both genes and samples. Another critical advantage is the efficient closed-form Gibbs sampling inference of the model parameters, which improves the computational complexity compared to the state-of-the-art methods. This is achieved by using a statistically well-founded data augmentation solution. In addition, GMNB  explicitly models the potential sequencing depth heterogeneity so that no heuristic preprocessing step is required. Experimental results on both synthetic and real-world RNA-seq data demonstrate the state-of-the-art performance of the GMNB method for temporal differential expression analysis of RNA-seq data. 

\section{Acknowledgments}

This research was supported in part by the NSF Awards CCF-1553281, CCF-1718513, and the USDA NIFA Award 06-505570-01006 to X. Qian and the NSF Award DBI-1532188 and the funding support from QNRF (NPRP9-001-2-001, NPRP7-1634-2-604), CSTR (2017-01), and CONACYT to P. de Figueiredo.

\begin{spacing}{1}
\bibliographystyle{abbrvnat} %

\bibliography{reference}
\end{spacing}
\end{document}